\documentclass[twocolumn,prl,showpacs]{revtex4} \usepackage{graphicx}
\usepackage{dcolumn}

\begin{document}
\title{On the Melting of Bosonic Stripes} \date{\today} \author{Guido
  Schmid$^{(1)}$ and Matthias Troyer$^{(1,2)}$} \affiliation{
  $^{(1)}$Theoretische Physik, Eidgen\"ossische Technische Hochschule
  Z\"urich, CH-8093 Z\"urich, Switzerland \\
  $^{(2)}$ Computation Laboratory, Eidgen\"ossische Technische Hochschule
  Z\"urich, CH-8092 Z\"urich, Switzerland}
\begin{abstract}
We use quantum Monte Carlo simulations to determine the finite temperature
phase diagram and to investigate the thermal and quantum melting of stripe
phases in a two-dimensional hard-core boson model.  At half filling and low
temperatures the stripes melt at a first order transition. In the doped
system, the melting transitions of the smectic phase at high temperatures
and the superfluid smectic (supersolid) phase at low temperatures are either
very weakly first order, or of second order with no clear indications for an intermediate
nematic phase.
\end{abstract}
 \maketitle
 
Stripe phases of lattice models with broken rotational and translational
symmetry can melt in two qualitatively different ways. One scenario is that
both symmetries can be restored at a single first or second order transition,
and the stripes melt directly into a normal fluid or superfluid phase. The
other scenario is that first the translational symmetry is restored when the
striped solid melts into a nematic (liquid crystal) phase with broken
rotational symmetry. The rotational symmetry is then restored in a second
melting transition of the nematic phase.

This quantum lattice problem shows similarities to the long standing problem
of the melting of a two dimensional crystal into a continuum model, where there
is also either a first order melting or the
Kosterlitz-Thouless-Halperin-Nelson-Young (KTHNY) scenario of
two Kosterlitz-Thouless transitions with an intervening hexactic phase~\cite{KTHNY}. Clear
 results for the classical version of this continuum problem were obtained
only recently in a simple model of hard disks
\cite{HardDisk}.
 
Current interest in quantum mechanical stripe phases and their melting stems
from the experimental observation of stripes in 
some high-$T_c$ superconductors \cite{Tranquada} and from the question
whether and how they are related to the occurrence of high temperature
superconductivity. Numerically, stripe phases have been found to be
competitive ground states of $t$-$J$-like models \cite{tsune,tjstripes}.
Analytically some theories of high temperature superconductivity are closely
linked to the existence of stripe and nematic phases \cite{Kivelson}. While
it is hard to study stripe phases in strongly correlated fermionic models,
because of the negative sign problem of quantum Monte Carlo, we can more accurately investigate
bosonic stripes using modern quantum Monte Carlo algorithms
\cite{loop,Sandvik}. Such bosonic models can appear as effective low
energy models neglecting nodal quasiparticles \cite{tsune,bosonmodels}. Like the cuprates, these 
bosonic models show competition and in some models coexistence of superfluidity and charge order.
 
In this Letter we focus on the simplest bosonic quantum model exhibiting
stripe order and determine its finite temperature phase diagram. We carefully
investigate its thermal and quantum melting transitions to address the
question of how stripe melting occurs in a quantum model of bosonic stripes.
The particular hard-core boson Hubbard Hamiltonian we study on a two-dimensional square lattice is
\begin{eqnarray}
H=-t\sum_{\langle
    {\bf i},{\bf j}\rangle}(a_{{\bf i}}^{\dagger}a_{{\bf j}}+ a_{{\bf j}}^{\dagger}a_{{\bf i}})
+V_2\sum_{\langle\langle {\bf i},{\bf j} \rangle\rangle} n_{{\bf i}} n_{{\bf j}}-\mu \sum_{\bf i} n_{\bf i},
\label{HCBham}
\end{eqnarray}
where $a_{\bf i}^{\dagger}$ $(a_{\bf i})$ is the creation (annihilation)
operator for hard-core bosons, $n_{\bf i}=a_{\bf i}^{\dagger} a_{\bf i}$ the
number operator, $V_2 \ge 0$ the next nearest neighbor Coulomb repulsions and
$\mu$ the chemical potential. The Hamiltonian (\ref{HCBham}) is equivalent to
an anisotropic spin-1/2 model with a nearest neighbor $XY$-like interaction
$J_{xy}=2t$ and next nearest neighbor Ising interaction $J'_{z}=V_2$ in a
magnetic field $h=2V_2-\mu$.

\begin{figure}
  \includegraphics[width=0.48\textwidth]{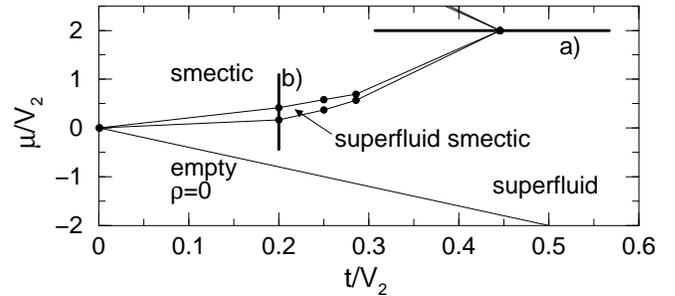}
\caption[]{
  Ground-state phase diagram of the hard-core boson Hubbard model Eq.
  (\ref{HCBham}) as function of $t/V_2$ and $\mu/V_2$. Due to particle-hole
  symmetry, the phase diagram is symmetric around the half filling line
  (density $\rho=1/2$ at $\mu=2V_2$) and the lower half is shown.  The thick
  lines $1)$ and $2)$ indicate the cuts along which we show the
  finite-temperature phase diagrams in Fig.~\ref{fig:Tc}. }

\label{fig:groundstate}
\end{figure}
In Fig.~\ref{fig:groundstate} we review the ground-state phase
diagram of this model which was previously studied by mean-field, renormalization group and local-update quantum Monte Carlo simulations \cite{pich,frey,batrouni3}.  At half filling (density $\rho=1/2$ at $\mu=2V_2$)
the ground state is a smectic for $t/V_2 < 0.446 \pm 0.006$ and the quantum
melting transition at low temperatures was found to be of first order (translation symmetry is broken only in one dimension, perpendicular to
the stripes, hence this phase is a smectic) .

Doping
this smectic stripe crystal a stable ``supersolid" phase with coexisting stripe
order and superfluidity was found, in
which the vacancies doped into the stripes form a superfluid.

\begin{figure}
  \includegraphics[width=0.48\textwidth]{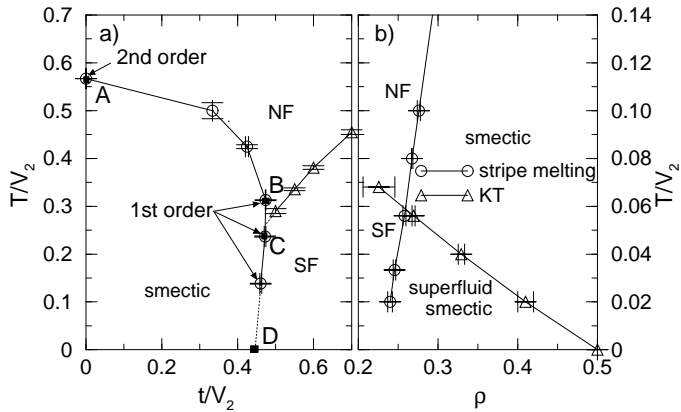}
\caption[]{
  (a) Finite temperature phase diagram as function of $t/V_2$ of the half
  filled model and (b) as function of the density $\rho$ along the respective
  lines of Fig.~\ref{fig:groundstate}. The normal fluid and superfluid phases
  are denoted by the symbols NF and SF.}
\label{fig:Tc}
\end{figure}
Before going into details we summarize our key results by presenting
the finite temperature phase diagrams along the lines indicated in the ground
state phase diagram Fig.~\ref{fig:groundstate}. At half filling (see Fig.
\ref{fig:Tc}a) we find that the transition between smectic and
superfluid is of first order at all temperatures.  At higher temperatures the
transition between smectic and normal fluid is of first order around
$t/V_2\approx 0.5$, but then becomes very weakly first order, before turning
second order as $t/V_2\rightarrow0$. In the doped system (see Fig.
\ref{fig:Tc}b) coexistence between the smectic phase and superfluidity
can be observed at low temperatures around half filling. Here the phase
transitions are all either very weakly first order or of second order.
Rotational and translational symmetry breaking occur, within the accuracy of
our simulations, at the same point.

To determine the finite-temperature phase diagram we have used quantum Monte
Carlo simulations, employing a directed loop quantum Monte Carlo algorithm in
the in the stochastic series expansion (SSE) representation
\cite{Sandvik}. The simulations were performed in the grand canonical
ensemble and do not suffer from any systematic errors apart from finite size
effects. In contrast to the loop algorithm \cite{loop} the directed loop
algorithm remains effective away from half filling where the loop algorithm
slows down exponentially.

We determine stripe order by measuring the order parameter
\begin{equation}
O_S = S_n(\pi,0)+S_n(0,\pi)
\end{equation}
where $S_n(k_x,k_y)$ is the charge structure factor at the wave vector
$(k_x,k_y)$. To investigate a nematic phase, we have to look for rotational
symmetry breaking in the kinetic energy or in the local charge correlations,
using as order parameters
\begin{eqnarray}
O_k=\frac{1}{V} \sum_{(x,y)} a_{(x,y)}^{\dagger} a_{(x+1,y)}^{\phantom{\dagger}} 
- a_{(x,y)}^{\dagger} a_{(x,y+1)}^{\phantom{\dagger}} + H.c.
\end{eqnarray}
or alternatively
\begin{eqnarray}
O_N = \sum_{(x,y)} n_{(x,y)}n_{(x+1,y)}-n_{(x,y)}n_{(x,y+1)}
\end{eqnarray}
where $n_{(x,y)}=a_{(x,y)}^{\dagger} a_{(x,y)}^{\phantom{\dagger}}$ is the
boson number operator at lattice site $(x,y)$ and $H.c.$ denotes the Hermitian
conjugate.

The extent of the superfluid phase is determined by measuring the superfluid
(number) density
\begin{eqnarray}
\rho_s = mT \langle W^2\rangle 
\label{eq:rhos}
\end{eqnarray}
where $W$ is the winding number in one of the directions and $m$ the 
mass of a boson. $\rho_s$ is finite in the superfluid phase and
jumps to zero from a universal value ${2\over \pi}mT_c$ at the Kosterlitz-Thouless transition.

\begin{figure}
  \includegraphics[width=0.4\textwidth]{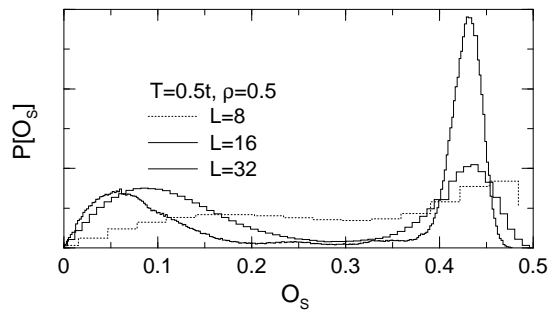}
\caption{
  Histograms of the stripe order parameter $O_S$ clearly show the
  double-peak structure of a first order transition which gets more pronounced
  as the system size is increased.  }
\label{fig:histo}
\end{figure}
We now discuss the phase diagrams and the nature of the phase transitions in
more detail, starting with the {\it half filled} system at $\mu=2V_2$, where a
first order quantum phase transition was found at $V_2/t = 2.24\pm 0.03$,
improving the previous estimate \cite{batrouni3}. When raising the temperature
we find that the transition remains of first order as the histograms for both
the stripe order parameter $O_S$ and the rotational symmetry breaking order
parameter $O_N$ show two clearly separated peaks corresponding to the two
coexisting phases (see Fig. \ref{fig:histo}). Careful finite size scaling of the
peak distances on system sizes $L=8$ up to $L=32$ show that the two peaks stay
separate as expected for a first order transition and do not merge as in a
second order one. For both order parameters we find the same phase transition
line, confirming a melting in a single strong first order transition up to at
least $T/t=2/3$ [from point (D) to (B) in Fig.~\ref{fig:Tc}a)].

To investigate whether the stripes melt into a superfluid or normal
fluid we measure the superfluid density $\rho_s$ at the coexistence line for
configurations in the fluid phase (determined by their value of $O_S$). At a
temperature $T/t=2/3$ [point (B)], $\rho_s$ drops below the universal jump 
value ${2\over \pi}mT$ as a function of system size $L$ and will thus scale to
zero in the infinite system. Hence the transition at this temperature is of
first order to a normal fluid phase in contrast to $T=0.5t$ [point (C)] where
$\rho_s$ is well above the universal value for all system sizes indicating a
direct first order transition between the smectic and a superfluid
phase~\cite{nelson}. There is thus a tricritical point between points $B$ and
$C$ where the Kosterlitz-Thouless phase transition of the superfluid becomes
first order.

At higher temperatures and larger repulsion $V_2$ it becomes harder to
determine the order of the phase transition. At $V_2=3t$ no double-peak structure
could be seen in the histograms for lattice sizes up to $L=32$ and the
transition is thus either second order or a very weak first order transition.
In the limit of infinite repulsion $V_2$ the lattice decouples into two
sub-lattices, each of which is equivalent to an Ising antiferromagnet with a
second order melting transition. We thus explore the possibility of a second
order transition.
\begin{figure}
  \includegraphics[width=0.48\textwidth]{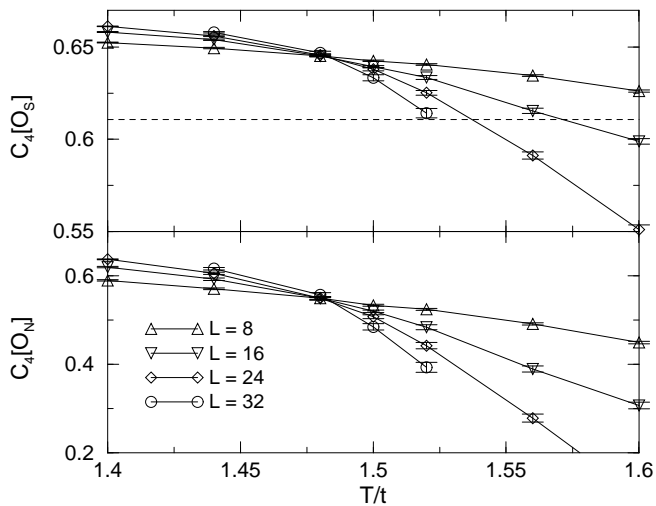}
\caption{
  Fourth order cumulants for $O_S$ and $O_N$ as function of the temperature
  ($V_1=0$, $V_2=3t$). The dashed horizontal line shows the critical value for a
  second order transition in the Ising universality class. The crossing point
  of the $4$-th order cumulant ratios for $O_S$ is not consistent with this
  value. }
\label{fig:binder}
\end{figure}
For a second order transition the $4$-th order cumulant ratios $C_{4}=1-\langle
O^4\rangle/3\langle O^2\rangle^2$ have a size independent value at the
transition point, which can be found as the crossing points of the cumulants
for different system sizes. Our results (see Fig.~\ref{fig:binder}) show that we indeed observe such crossings, which is and indication for second order transition. Both the
$4$-th order cumulant ratios for both $O_N$ and $O_S$ cross at the same
temperature within the accuracy of our results, indicating that
translational symmetry breaking (measured by $O_S$) and rotational symmetry
breaking (measured by $O_N$) happen at the same or at very close temperatures.

The phase diagram of the {\it doped system} away from half filling shows an
additional ``supersolid'' phase where the vacancies doped into the stripes
form a superfluid smectic with broken translational and rotational symmetry as
well as a finite superfluid density. 
Since the rotational symmetry is spontaneously broken in the smectic phase,
the superfluid becomes anisotropic and Eq. (\ref{eq:rhos}) for the superfluid
density needs to be modified. The geometric mean of the superfluid densities
perpendicular and parallel to the stripes $\rho_s=\sqrt{\rho_{s,||}
\cdot \rho_{s,\perp}}$ is the quantity exhibiting the universal jump at the
Kosterlitz-Thouless transition and was used in the determination of the phase
transition. Care was taken that the anisotropy did not become too large to
introduce systematic errors as pointed out by Prokof'ev and Svistunov
\cite{rhos}. The superfluid densities parallel ($\rho_{s,||}$) and
perpendicular ($\rho_{s,\perp}$) to the stripe order were measured for each
configuration from the winding numbers in either $x$ or $y$ direction
depending on whether $S_n(0,\pi)$ was larger or smaller than $S_n(\pi,0)$.

Although superfluidity coexists with
smectic order, it is strongly suppressed and $T_c$ and $\rho_s$ vanish
linearly as half filling is approached. Using $m_{||}=1/2t$ and 
$m_{\perp}=\frac{1}{2t^2/(4V_2)}$ as the boson masses parallel 
and perpendicular to the stripes, respectively, we obtain 
$\rho_s= (0.78 \pm 0.07)|\rho-\frac{1}{2}|$ consistent with the value obtained 
previously for a dilute gas of bosons \cite{bernardet}.

\begin{figure}
  \includegraphics[width=0.48\textwidth]{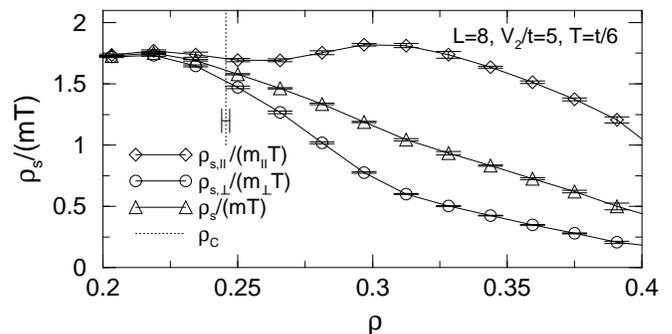}
\caption{
  Superfluid density parallel ($\rho_{s,||}$) and perpendicular ($
  \rho_{s,\perp} $) to the stripes for $L=8$, $V_2/t=5$ and $T=t/6$.}
\label{fig:dw0.1666}
\end{figure}
In order to investigate the nature of the doping-driven melting of the smectic
phase we repeat a similar procedure as at half filling. In contrast to the
half filled case, with a clear jump in the values of $O_S$ and $O_N$ at the
phase transition, or in the doped case of the nearest neighbor model with a
jump in density $\rho$ at the doping-driven melting transition \cite{schmid},
no such jump or double-peak structure was observed here. We checked
the order parameters $O_S$, $O_N$, $O_k$, and the density $\rho$ in systems with size up to $L=48$. Repeating the simulations of Ref.~\cite{batrouni3}, where a local update algorithm was used instead of our
nonlocal algorithm we cannot reproduce the jump in the superfluid
density seen there. Instead we find the smooth behavior shown in Fig.
\ref{fig:dw0.1666} for the same system, which is also in agreement with analytical
considerations using mean-field and spin-wave analysis~\cite{pich} or
renormalization group calculations~\cite{frey}.  Since our results are well
converged and do not suffer from any systematic error, we believe the
differences to be due to the method used to calculate
the superfluid densities in Ref. \cite{batrouni3}.

At $T=t/6$ we checked whether rotational and translational symmetry breaking
occur at the same critical density by comparing estimates obtained using
Binder cumulant ratios for $O_S$ and $O_N$ as well as a determination of the
critical point from the maxima of the susceptibility associated with $O_k$.
The estimates agree within the error bars: $\rho_c=0.2468 \pm 0.0025$ at
$T=t/6$.

Finally we want to discuss our results in view of the three possibilities for 
the nature of the melting transitions of i) a first order transition,
ii) a single second order transition or iii) two separate second order
transitions. Close to half filling and at low temperatures we have very clear
evidence for a single first order transition. The results for thermal melting
at higher temperatures and for the doping-driven melting at low temperatures
are not as clear-cut. While a very weak first order transition is always
possible, it is more likely that we have second order transitions, especially
in view of the fact that in the limit $t/V_2\rightarrow\infty$ there will be a
second order transition in the Ising universality class. One way to
distinguish between the scenarios (ii) of one single transition or (iii) of
two second order transitions with an intervening nematic phase is that in the
latter case both transitions should be in the Ising universality class since
they each break a $Z_2$ symmetry. At the transition we would thus expect that
the Binder cumulant ratios at $T_c$ takes on the value $C_4=0.6106900(1)$
~\cite{kamieniarz}. The fact that our results for the cumulant ratio of $O_S$
are close but not identical to this universal value for an Ising transition
rather indicates secnario (ii), a single second order transition in a
different universality class. In the doped system our data both for the cumulant ratios and the scaling of the susceptibility of $O_k$ are actually closer to the three dimensioonal $XY$ universality class predicted by Ref. \cite{frey} than the Ising universality class. More extensive simulations on larger systems, using newly developed flat histogram methods for quantum systems \cite{QWL} will be needed to more accurately determine the universality class of this phase transition.

While the exact order and universality class of the melting transition are hard to determine
 -- which is not
surprising given the difficulties known from the similar classical problem in
the continuum -- our results are interesting also with respect to the
existence of a nematic phase. Such a phase must be restricted to a
very narrow temperature and doping regime, smaller than the resolution of our
simulations. It might be possible to stabilize a nematic phase
with additional terms in the Hamiltonoan. One suggestion
\cite{kivelson} is to add a term proportional to the square of the nermatic
order parameter $-V O_k^2$, to stabilize a nematic phase. This term gives two
contributions: a nearest neighbor repulsion $2V\sum_{\langle {\bf i},{\bf
    j}\rangle }n_{\bf i} n_{\bf j}$ and an additional next nearest neighbor
hopping term $V\sum_{\langle\langle i,j \rangle\rangle}(a^{\dag}_{\bf i}a_{\bf
  j}+ a_{\bf j}^{\dag}a_{{\bf i}})$.  These frustrating terms, the second of
which unfortunately causes a negative sign problem for quantum Monte Carlo
simulations, might be important for the stability of an extended nematic
phase. It is interesting to compare this model with the related frustrated
Heisenberg model on a square lattice which however exhibits translational and
rotational symmetry breaking only in the ground state at $T=0$ \cite{singh}.

We are grateful for stimulating discussions with G. Batrouni, S. Kivelson and
S. Todo and acknowledge support of the Swiss National Science
Foundation.  The simulations were performed on the Asgard Beowulf cluster at ETH
Z\"urich, using a parallelizing C++ library for Monte Carlo simulations
\cite{alea}.

\end{document}